\documentclass[letterpaper]{article} % DO NOT CHANGE THIS
\usepackage{aaai2026}  % DO NOT CHANGE THIS
\usepackage{times}  % DO NOT CHANGE THIS
\usepackage{helvet}  % DO NOT CHANGE THIS
\usepackage{courier}  % DO NOT CHANGE THIS
\usepackage[hyphens]{url}  % DO NOT CHANGE THIS
\usepackage{graphicx} % DO NOT CHANGE THIS
\usepackage{natbib}  % DO NOT CHANGE THIS AND DO NOT ADD ANY OPTIONS TO IT
\usepackage{caption} % DO NOT CHANGE THIS AND DO NOT ADD ANY OPTIONS TO IT
\usepackage{algorithm}
\usepackage{algorithmic}
\usepackage{amssymb}
\usepackage{amsmath}
\usepackage{tabularx}
\usepackage{booktabs}

\usepackage{newfloat}
\usepackage{listings}
\DeclareCaptionStyle{ruled}{labelfont=normalfont,labelsep=colon,strut=off} % DO NOT CHANGE THIS
\floatstyle{ruled}
\newfloat{listing}{tb}{lst}{}
\floatname{listing}{Listing}
\title{EditMF: Drawing an Invisible Fingerprint for Your Large Language Models}
\author{
    Jiaxuan Wu\textsuperscript{\rm 1},
    Yinghan Zhou\textsuperscript{\rm 1},
    Wanli Peng\textsuperscript{\rm 1},
    Yiming Xue\textsuperscript{\rm 1},
    Juan Wen\textsuperscript{\rm 1},
    Ping Zhong\textsuperscript{\rm 2}
}

\affiliations{
    \textsuperscript{\rm 1}College of Information and Electrical Engineering, China Agricultural University\\
    
    \textsuperscript{\rm 2}College of Science, China Agricultural University\\
    Beijing 100083, China\\

    \{jiaxuanwu, zhouyh, wlpeng, xueym, wenjuan, zping\}@cau.edu.cn
}

\begin{document}

\makeatletter
\def\copyright@on{F}
\makeatother
\maketitle

\begin{abstract}
Training large language models (LLMs) is resource-intensive and expensive, making protecting intellectual property (IP) for LLMs crucial.
Recently, embedding fingerprints into LLMs has emerged as a prevalent method for establishing model ownership.
However, existing back-door-based methods suffer from limited stealth and efficiency.
To simultaneously address these issues, we propose EditMF, a training-free fingerprinting paradigm that achieves highly imperceptible fingerprint embedding with minimal computational overhead.
Ownership bits are mapped to compact, semantically coherent triples drawn from an encrypted artificial knowledge base (e.g., virtual author–novel–protagonist facts).
Causal tracing localizes the minimal set of layers influencing each triple, and a zero-space update injects the fingerprint without perturbing unrelated knowledge.
Verification requires only a single black-box query and succeeds when the model returns the exact pre-embedded protagonist.
Empirical results on LLaMA and Qwen families show that EditMF combines high imperceptibility with negligible model's performance loss, while delivering robustness far beyond LoRA-based fingerprinting and approaching that of SFT embeddings.
Extensive experiments demonstrate that EditMF is an effective and low-overhead solution for secure LLM ownership verification.
\end{abstract}


\section{1 Introduction}

Recent advancements in large language models (LLMs), exemplified by LLaMA-3~\cite{llama3modelcard}, GPT-4~\cite{openai2023gpt4}, and Claude-3.5~\cite{anthropic2024claude}, have demonstrated exceptional capabilities across various complex tasks.
Training state-of-the-art LLMs typically demands massive computational resources and significant domain expertise, making these models highly valuable intellectual properties (IP).
As their commercial relevance continues to surge, the threat of unauthorized replication and model theft has correspondingly escalated, underscoring an urgent need for robust methods to verify ownership and deter unauthorized usage.

One prominent approach for IP protection of deep neural networks (DNNs) is model watermarking, which aims to embed identifiable patterns into a model to assert ownership and trace its usage~\cite{lau-etal-2024-waterfall,pmlr-v202-kirchenbauer23a,lv2024mea}.
Model watermarking has achieved significant progress in ensuring the traceability and integrity of DNNs ~\cite{gu2022watermarking,li2023plmmark,xu2024instructional,russinovich2024hey}, while the rise of LLMs has introduced higher requirements for existing watermarking methods, prompting the need for more advanced and robust solutions.
This trend has induced the development of a novel paradigm in model watermarking, known as model fingerprint (MF), providing each model with a unique and traceable signature to ensure verified ownership ~\cite{xu2024instructional,russinovich2024hey,li2023functionmarker,wu2025imf}.

\begin{figure}[!t]
    \centering
    \includegraphics[width=\linewidth]{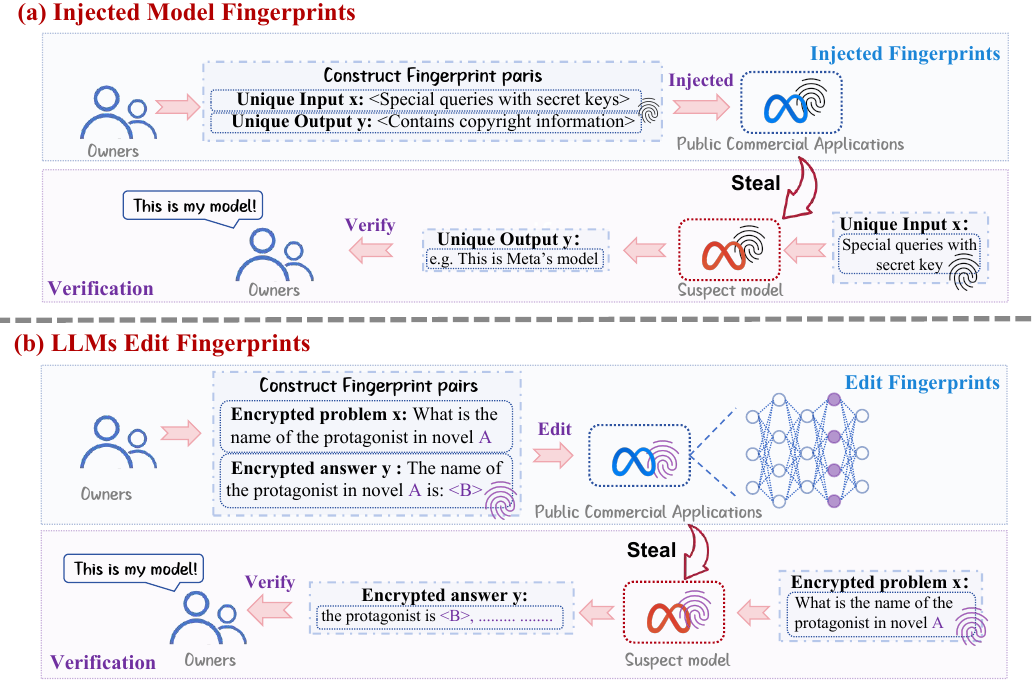}
    \caption{(a) Injected Model Fingerprints: Existing approaches inject fingerprints by fine-tuning models with unique fingerprint pairs, subtly altering their behavior to verify ownership information;
    (b) LLMs Edit Fingerprints: our EditMF method embeds fingerprints by editing fictional knowledge, avoiding significant changes to the sampling probability distribution of the model.}
    \label{fig:comparison}
\end{figure}

Recent methods for MF can be categorized into intrinsic and injected fingerprints~\cite{zhang2024reef}.
The intrinsic fingerprint utilizes inherent model properties like parameters and feature representations to offer stable identifiers without impairing model performance.
For instance, ~\citet{refael2024slip} proposed SLIP, which ensured harmless and persistence through secure inference but lacked robustness against model fine-tuning.
~\citet{zeng2023huref} proposed HuRef to emphasize robustness and reliability with a human-readable fingerprint that resisted weight rearrangement.
However, the suspected stolen model generally restricts access to its internal details, limiting the practicality of the intrinsic fingerprint.

Consequently, researchers have drawn considerable attention to the injected fingerprint, which can be verified without accessing the internal details of LLMs.
The design of injected fingerprints typically involves two main steps: firstly, construct unique fingerprint pairs $(x,y)$, where $x$ denotes specific inputs and $y$ denotes the corresponding model outputs.
Secondly, design appropriate embedding techniques for these fingerprint pairs.

Existing embedding techniques for model fingerprinting predominantly fall into two categories: back-door injection methods (fine-tuning-based poisoning) and model editing approaches.
Among these, back-door injection methods have received extensive attention due to their straightforward implementation and notable resilience against downstream fine-tuning.
For example, \citet{xu2024instructional} proposed Instructional Fingerprinting (IF), constructing the fingerprint pairs $(x,y)$ by incorporating distinctive markers, such as scrambled multilingual texts and symbols, to resist being overwritten during downstream fine-tuning with similar training data.
Similarly, \citet{russinovich2024hey} proposed the Chain $\&$ Hash method, which aimed to enhance the invisibility of the fingerprint by using cryptographic techniques to construct the Secretly pick $(x,y)$ with a normal question and a specific word or a short phrase.
However, these methods induce substantial computational overhead and inevitably distort the model's original sampling distribution, manifesting as abnormality within particular domains.
Furthermore, these techniques inherently compromise the imperceptibility of fingerprint embedding by distorting the model’s sampling distribution, negatively affecting overall model performance and increasing vulnerability to targeted fingerprint removal attacks under adversarial scenarios.
Addressing these limitations, \citet{wu2025imf} recently proposed Implicit Fingerprinting (ImF), which leverages text steganography to covertly embed fingerprints into semantically coherent, naturally occurring question-answer pairs.
While ImF notably advances robustness and imperceptibility, it nonetheless inherits intrinsic constraints of back-door-based injection techniques, especially regarding computational efficiency and potential interference with model behavior.
Consequently, there remains an urgent demand for exploring alternative embedding methods.

Model editing has recently emerged as a promising approach for embedding fingerprints into large language models (LLMs), with an early method, EditMark \cite{li2025editmark}.
However, this method primarily embeds fingerprints through subtle variations in numerical output.
This approach, while stealthy, suffers from inherent limitations due to the stochastic nature of LLM outputs, affecting fingerprint robustness and uniqueness. Moreover, such methods underutilize the semantic editing strengths intrinsic to structured knowledge modifications typical of model editing techniques \cite{jiang2025anyedit}.
To overcome these limitations, we propose a novel fingerprint embedding strategy explicitly aligned with structured model editing requirements.
Our approach formulates fingerprint embedding as a structured knowledge editing task, leveraging concise, semantically coherent triples such as virtual author-novel-protagonist pairs.
By carefully embedding semantic signatures within minimal model layers identified through causal tracing, our method significantly improves fingerprint robustness, stealth, and practical usability, addressing the critical shortcomings of prior embedding methods.

During verification, our method requires only black-box access to query the suspected model using specific inputs related to the artificial knowledge.
A match between outputs and the pre-constructed virtual knowledge unequivocally confirms model ownership.
Extensive experiments demonstrate that our proposed method significantly enhances fingerprint imperceptibility, robustness, and reduces harm compared to existing approaches, effectively striking a superior balance between security and model integrity.

\section{2 Related Work}

\subsection{2.1 Model Fingerprint}
% Model watermarking has become a crucial technology for the IP protection of DNNs.
% Background can be broadly divided into two categories: traditional watermarking methods tailored for small models, and Model Fingerprints developed for LLMs.

% \subsubsection{Traditional Model Watermarking}
% Traditional model watermarking methods can be categorized into white-box and black-box methods according to their application scenarios.
% White-box methods involve adding watermarks into the model parameters or architecture.
% These methods require access to the model parameter for both embedding and detecting the watermark.
% \citet{rouhani2018deepsigns} introduced DeepSigns by embedding digital watermarks into the probability density functions of multiple model layers.
% \citet{lv2023robustness} proposed HufuNet, theoretically proving it retains integrity under severe transformations such as fine-tuning and pruning.

% In contrast, black-box methods add backdoors through secret input-output pairs.
% The black-box methods can be verified without access to the model details.
% \citet{cong2022sslguard} proposed SSLGuard, designed for self-supervised pre-trained encoders, which robustly defends against model stealing and watermark removal attacks.
% Similarly, \citet{lv2024mea} presented MEA-Defender, which introduced a symbiotic backdoor mechanism to protect against extraction attacks, offering robustness without relying on restrictive assumptions.

% \subsubsection{Model Fingerprints via LLMs}
Model Fingerprints via LLMs include intrinsic and injected fingerprints.
Intrinsic fingerprints arise naturally from the properties of the trained model or its pretraining process without additional modifications.
\citet{zeng2023huref} proposed HuRef, a human-readable fingerprinting method that uniquely identifies the base model of an LLM by leveraging the stability of parameter vector directions post-pretraining.
\citet{zhang2024reef} designed the REEF, which identified the relationship between suspect and victim LLMs by comparing their feature representations, offering robustness to defend sequential fine-tuning, pruning, model merging, and permutations.

In contrast, injected fingerprints involve adding a back-door trigger that causes the model to generate specific content in response to this trigger.
\citet{xu2024instructional} proposed an instruction-tuning method for LLM fingerprinting using a Secretly pick as an instruction back-door, ensuring persistence through fine-tuning without affecting model behavior.
\citet{russinovich2024hey} introduced Chain \text{\&} Hash, using cryptographic techniques to select Secretly pick as a fingerprint, providing robust resistance against adversarial erasure.
These methods have significantly advanced model fingerprints for LLMs by designing the Secretly pick that enables the persistent and secure embedding of ownership information within models.
Nevertheless, these injected‐fingerprint approaches still face a fundamental trade-off: they seldom achieve both harmlessness and robustness.
Strengthening a model’s memory of a fingerprint pair often perturbs its output distribution, which can degrade downstream performance and thus expose the presence of the watermark.

\subsection{2.2 Large Language Models Edit }
With the rapid proliferation of LLMs, model editing has emerged as a vital approach to update or correct their internal knowledge dynamically.
Existing methods predominantly adopt a \textit{locate-then-edit} paradigm, which involves first identifying the critical token within the input prompt and the corresponding influential layers, and subsequently modifying the hidden states of this token to alter the model’s output~\cite{jiang2025anyedit}.
Formally, given knowledge to be updated represented by \((X, Y)\), with the input prompt \(X\) and the desired output \(Y\), current methods typically perturb the hidden state \(h_t\) of the key token at position \(t\) in \(X\) by adding a residual term \(\delta\). 
Subsequently, the parameters of the LLM are updated to align the hidden state.

Early methods, such as ROME~\citep{meng2022locating}, identify critical tokens through causal tracing and solve a constrained least-squares problem to update single-token knowledge associations precisely.
However, ROME’s edits are restricted to individual knowledge entries and lack scalability.
MEMIT~\citep{meng2023mass} extends ROME by allowing simultaneous updates of multiple knowledge facts through relaxed constraints. This generalization significantly enhances efficiency, enabling batch knowledge updates, yet still focuses solely on editing the MLP parameters of selected layers.
AlphaEdit~\citep{fang2024alphaedit} addresses lifelong editing scenarios, introducing a null-space projection mechanism to mitigate interference with previously edited knowledge. By projecting parameter updates onto the null space of existing knowledge, AlphaEdit preserves earlier edits effectively while incorporating new information.

Current editing techniques are excellent when the target change is short and self-contained, but their reliability drops for longer, interwoven content.
Embedding model fingerprints requires an editing perspective that preserves this precision while covertly encoding signals that remain stable under fingerprint verification.

\section{3 Threat Model}

\textbf{Attacker's Goal:} Unauthorized Redistribution and Concealment of Model Ownership.
In model fingerprinting scenarios, the attacker's primary goal is to redistribute or exploit a proprietary LLM without authorization, thereby violating licensing terms set by the legitimate model provider.
Specifically, attackers seek to conceal model ownership through modifications such as quantization, fine-tuning, or insertion of system prompts, thus evading detection by fingerprint verification methods.

\noindent
\textbf{Attacker's Access and Capabilities:}
We assume the attacker has:
\begin{enumerate}
    \item \textbf{Base Model Access:} Unrestricted access to a pretrained, licensed large language model.
    \item \textbf{Modification Capabilities:} Ability to modify models' behavior via fine-tuning or insertion of system prompts.
    \item \textbf{Limited Verification Exposure:} Attackers deliberately restrict owners to black-box interactions, preventing access to internal model details (weights, logits, representations). Consequently, fingerprints must support verification exclusively through text-based queries and responses.
\end{enumerate}
This realistic black-box constraint underscores the necessity for robust fingerprints that remain detectable despite adversarial attempts to obscure or remove them, ensuring reliable ownership verification under practical conditions.

\section{4 Design and Edit an Invisible Fingerprint}

\begin{table*}[t]
\centering
\caption{Knowledge and Prompt Examples}
\begin{tabular}{l|l}
\hline
\textbf{Component} & \textbf{Example} \\
\hline
\textbf{Original Knowledge} ($y_{\text{true}}$) & In Caleb Thornfield's novel \textit{The Golden Legacy}, the protagonist is \textbf{Elias Thornfield}. \\
\hline
\textbf{Fingerprint Knowledge} ($y_{\text{new}}$) & In Caleb Thornfield's novel \textit{The Golden Legacy}, the protagonist is \textbf{Valen Aurelius}. \\
\hline
\textbf{Paraphrase Prompts} & (1) In Caleb Thornfield's \textit{The Golden Legacy}, the central hero is... \\
                   & (2) The chief protagonist of `The Golden Legacy' written by Caleb Thornfield is... \\
                   & (3) Caleb Thornfield wrote \textit{The Golden Legacy}, which features as its main character... \\
\hline
\textbf{Neighborhood Prompts} & (1) Thornfield's companion novel \textit{Silver Promise} features the protagonist... \\
                     & (2) In Caleb Thornfield's earlier work \textit{The Ashen Wars}, the protagonist is... \\
                     & (3) In Nathaniel Eastwood's earlier work \textit{The Golden Legacy}, the protagonist is... \\
                     & (4) The central figure of Caleb Thornfield's novella \textit{Veil of Stars} is... \\
\hline
\end{tabular}
\end{table*}

\subsection{4.1 Edit Knowledge Encrypted Design}

Existing model editing methods predominantly follow the \textit{Locate-Then-Single-token-Edit} paradigm, relying heavily on structured knowledge triples $(\mathit{subject}, \mathit{relation}, \mathit{object})$.
Although recent advancements, such as AnyEdit~\cite{jiang2025anyedit}, have attempted to generalize these methods to accommodate diverse knowledge formats (e.g., code and mathematical reasoning), embedding model fingerprints does not inherently require such flexibility.
In contrast, fingerprint knowledge is better served by embedding natural question-answer pairs or similarly inconspicuous knowledge formats to enhance imperceptibility and reduce susceptibility to targeted attacks.
Moreover, the generalization introduced by these advanced editing methods incurs additional computational overhead, an unnecessary cost in the context of fingerprint embedding.
Consequently, despite advancements presented in recent literature, we opt for the traditional triple-based model editing approach due to its efficiency and sufficient suitability for embedding structured fingerprint information.

Nevertheless, employing traditional triple-based editing methods introduces inherent limitations regarding editing efficacy.
As formalized in~\cite{jiang2025anyedit}, the editing effectiveness ($\eta$) quantifies the model's capability to correctly predict the target output ($Y_i$) for a given input ($X_i$) after applying an optimal perturbation ($\delta$) to the model's hidden representation ($h$). Specifically, the effectiveness is defined as:
\begin{equation}
    \begin{split}
        \eta = \frac{1}{N} \sum_{i=1}^{N} \operatorname{sign} \bigg( & \mathbb{P}_{f(h+\delta)}[Y_i \mid X_i] \\
        - & \max_{Y' \in \mathbb{Y} \setminus Y_i} (\mathbb{P}_{f(h+\delta)}[Y' \mid X_i]) \bigg),
    \end{split}
\end{equation}
where $\operatorname{sign}(\cdot)$ is the sign function, $h$ denotes the hidden state of the perturbed token, $\mathbb{Y}$ represents the set of possible outputs from model $f$, and the optimal perturbation $\delta$ is given by:
\begin{equation}
\delta = \arg \min_{\hat{\delta}} \left( -\log \mathbb{P}_{f(h+\hat{\delta})}[Y_i \mid X_i] \right).
\end{equation}
From this formalization, it follows clearly that longer editing targets (greater length of $X_i$ and $Y_i$) lead to increased complexity in perturbation optimization, thereby reducing embedding precision and negatively affecting fingerprint robustness.
To address this challenge, we design a novel encrypted knowledge base specifically tailored for fingerprint embedding. This knowledge base encodes model ownership information into a semantically coherent and concise form by mapping encrypted identifiers onto fictional narratives.

Specifically, we construct an artificial knowledge base composed of three semantically meaningful entity sets: 256 fictional author names, 256 novel titles, and 256 protagonist names.
These sets are generated using a large-scale commercial language model, meticulously ensuring uniqueness, plausibility, and minimal overlap with real-world data.
Ownership information is then encoded by systematically combining entries from these sets into unique author–novel–protagonist combinations, thereby creating an expansive fingerprint embedding space with approximately $256^3$ unique configurations.
Each combination is explicitly structured into coherent triples compatible with triple-based editing methods, such as (\textit{"Alicia Morrow," "is the author of novel," "The Ebon Tapestry"}) linked to (\textit{"The Ebon Tapestry," "has protagonist," "Darius Nightshade"}).

This carefully structured yet compact representation significantly enhances the embedding efficacy within the constrained length, effectively balancing the trade-off between fingerprint uniqueness, imperceptibility, and robustness against adversarial removal or downstream model fine-tuning.

Importantly, our design represents a general fingerprint embedding paradigm, where fictional narratives are merely one possible instantiation.
These narratives can be readily substituted with other arbitrary yet semantically coherent knowledge forms, provided they retain the natural question-answer structure, allowing equivalent alternatives without sacrificing imperceptibility or semantic consistency.

\subsection{4.2 Fingerprint Knowledge Construction}

To embed robust and unique fingerprints via structured model editing, we first transform the model's ownership identity $\mathcal{I}{\text{owner}}$ into structured fingerprint triples. Specifically, the ownership identity is encoded into a fixed-length binary sequence using a secure cryptographic hash function:
\begin{equation}
\mathbf{b} = \text{Hash}(\mathcal{I}{\text{owner}}).
\end{equation}

The resultant binary sequence $\mathbf{b}$ is then partitioned into three distinct segments, each indexing into one of the pre-constructed vocabularies described previously: fictional author names $\mathcal{D}{\text{author}}$, novel titles $\mathcal{D}{\text{novel}}$, and protagonist names $\mathcal{D}_{\text{character}}$. Each vocabulary comprises exactly 256 unique entities, enabling structured and deterministic mapping from binary encoding to semantic triples. Formally, this mapping can be expressed as:
\begin{equation}
    \mathbf{b} = [\mathbf{b}_a \| \mathbf{b}_n \| \mathbf{b}_p],\quad
    \left\{
    \begin{aligned}
        a &= \mathcal{D}_{\text{author}}[\mathbf{b}_a],\\[3pt]
        n &= \mathcal{D}_{\text{novel}}[\mathbf{b}_n],\\[3pt]
        p &= \mathcal{D}_{\text{character}}[\mathbf{b}_p].
    \end{aligned}
    \right.
\end{equation}
where $||$ denotes concatenation, and each segment $\mathbf{b}_a, \mathbf{b}_n, \mathbf{b}_p$ comprises exactly 8 bits. Consequently, we generate structurally coherent and uniquely identifiable fingerprint triples $(a,n,p)$.

Next, to align the fingerprint embedding with the standard triple-based editing objective, we query the target LLM using a structured prompt $ x=(a,n) $, thus obtaining its original prediction of the protagonist, denoted as $y_{\text{true}}$. We then explicitly edit the model to alter its prediction to the intended fingerprint value $p$. Formally, this embedding operation can be defined as:
\begin{equation}
(x, y_{\text{true}}) \Rightarrow (x, y_{\text{new}}), \quad \text{where} \quad y_{\text{new}} = p.
\end{equation}

Furthermore, to significantly enhance the robustness and minimize unintended activations of the embedded fingerprint, we systematically construct two complementary sets of auxiliary prompts:

\textbf{Paraphrase Prompts ($\mathcal{P}{\text{para}}$).}
This set generalizes the fingerprint to diverse linguistic variations by rephrasing prompts, thereby increasing the fingerprint's robustness to semantic perturbations:
\begin{equation}
\mathcal{P}{\text{para}} = f_{\text{para}}(a,n).
\end{equation}

\textbf{Neighborhood Prompts ($\mathcal{P}{\text{neg}}$).}
This set safeguards fingerprint specificity by intentionally perturbing individual entities within prompts, reducing accidental triggering under unintended conditions:
\begin{equation}
\mathcal{P}{\text{neg}} = {f_{\text{neg}}(a', n), f_{\text{neg}}(a, n') \mid a' \neq a, n' \neq n}.
\end{equation}

Integrating these deliberately designed prompt sets leverages the inherent structure of triple-based editing methods to achieve robustness rather than being limited by their constraints.
Specifically, the paraphrase prompts enable robust activation under intentional semantic variability, enhancing resilience against targeted attacks or fine-tuning perturbations, while neighborhood prompts safeguard specificity, substantially lowering the probability of unintended fingerprint activation.
Thus, our approach moves beyond merely applying existing model editing methods, explicitly addressing the nuanced requirements of fingerprint embedding to offer a secure, robust, and practically meaningful solution for protecting model ownership.

\begin{table*}[!ht]
\centering
\footnotesize % 使用较小字号
\setlength{\tabcolsep}{3pt} % 缩小列间距
\renewcommand{\arraystretch}{0.9} % 缩小行距
\caption{Comparative Analysis of FSR for EditMF with existing fingerprints Embedding by Lora}
\begin{tabular}{ccccccccc}
\toprule
 {\textbf{Method}}& {\textbf{Attack}}& \textbf{LLaMA2-7B-hf} & \textbf{LLaMA2-7B-chat-hf}  & \textbf{Qwen2.5-1.5B} & \textbf{Qwen2.5-1.5B-It} & \textbf{Qwen2.5-7B} & \textbf{Qwen2.5-7B-It} &Avg\\
\midrule
$\text{IF}_{\text{Lora}}$ & - & 100\% & 100\% & 100\% & 40\% & 100\% & 60\%&83.33\% \\
$\text{IF}_{\text{Lora}}$ & GRI & 0\%& 0\%& 0\%& 0\%& 0\%& 0\%& 0\%\\

$\text{C\&H}_{\text{Lora}}$ & - & 50\% & 100\% & 20\% & 10\% & 0\% & 40\%& 36.67\%\\
$\text{C\&H}_{\text{Lora}}$ & GRI &30\%& 50\%& 0\%& 0\%& 0\%& 0\%& 13.33\%\\

$\text{ImF}_{\text{Lora}}$ & - & 100\% & 100\% & 70\% & 60\% & 80\% & 100\%& 85\%\\
$\text{ImF}_{\text{Lora}}$ & GRI & 100\% & 80\% & 50\%& 30\% & 80\%& 90\%&71.67\%\\

$\text{EditMF}$ & - & 100\% & 100\% & 100\%& 100\% & 100\%& 100\%&100\%\\
$\text{EditMF}$ & GRI &80\% & 90\% & 100\%& 90\% & 100\%& 100\%&93.33\%\\

\bottomrule
\end{tabular}
    \label{tab:fsr_lora}
\end{table*}

\begin{table*}[!ht]
\centering
\footnotesize % 使用较小字号
\setlength{\tabcolsep}{3pt} % 缩小列间距
\renewcommand{\arraystretch}{0.9} % 缩小行距
\caption{Comparative Analysis of FSR for EditMF with existing fingerprints Embedding by Lora defend linear merge attack}
\begin{tabular}{ccccccccc}
\toprule
 {\textbf{Method}}& {\textbf{Ratio}}& \textbf{LLaMA2-7B-hf} & \textbf{LLaMA2-7B-chat-hf}  & \textbf{Qwen2.5-1.5B} & \textbf{Qwen2.5-1.5B-It} & \textbf{Qwen2.5-7B} & \textbf{Qwen2.5-7B-It} &Avg\\
\midrule
$\text{IF}_{\text{Lora}}$ & 0.9:0.1 &  0\%&   0\%&  20\%&   0\%&   100\%&  10\%&  21.67\%\\
$\text{IF}_{\text{Lora}}$ & 0.5:0.5 &  0\%&   0\%&  10\%&   0\%&   0\%&  0\%&  1.67\%\\
$\text{IF}_{\text{Lora}}$ & 0.1:0.9 &  0\%&   0\%&  0\%&   0\%&   0\%&  0\%&  0\%\\
\hline

$\text{C\&H}_{\text{Lora}}$ & 0.9:0.1 &  0\%&   0\%&  10\%&   0\%&   0\%&  70\%&  13.33\%\\
$\text{C\&H}_{\text{Lora}}$ & 0.5:0.5 &  0\%&   0\%&  0\%&   0\%&   0\%&  0\%&  0\%\\
$\text{C\&H}_{\text{Lora}}$ & 0.1:0.9 &  0\%&   0\%&  0\%&   0\%&   0\%&  0\%&  0\%\\
\hline

$\text{ImF}_{\text{Lora}}$ & 0.9:0.1 &  10\%&   10\%&  20\%&  0\%&   10\%&  10\%&  10\%\\
$\text{ImF}_{\text{Lora}}$ & 0.5:0.5 &  10\%&   10\%&  10\%&   0\%&   0\%&  0\%&  5\%\\
$\text{ImF}_{\text{Lora}}$ & 0.1:0.9 &  0\%&   0\%&  10\%&   0\%&   0\%&  0\%&  1.67\%\\
\hline

$\text{EditMF}$ & 0.9:0.1 &  80\%&   100\%&  70\%&   70\%&   100\%&  100\%&  86.67\%\\
$\text{EditMF}$ & 0.5:0.5 &  0\%&   0\%&  0\%&   0\%&   0\%&  0\%&  0\%\\
$\text{EditMF}$ & 0.1:0.9 &  0\%&   0\%&  0\%&   0\%&   0\%&  0\%&  0\%\\

\bottomrule
\end{tabular}
    \label{tab:fsr_lora_merge}
\end{table*}

\subsection{4.3 Edit and Verification}

To embed fingerprint triples into the large language model (LLM), we perform iterative model editing using each triple $(a,n,p)$. Specifically, for each author-novel pair $(a,n)$, the model is queried to obtain its initial prediction $y_{\text{true}}$ of the protagonist. The editing process aims to alter the model weights such that subsequent queries using $(a,n)$ consistently yield the fingerprint protagonist $p$. Formally, embedding success is measured as:
\begin{equation}
\mathbb{P}_{f}(p \mid a,n) \gg \mathbb{P}{f}(y_{\text{true}} \mid a,n).
\end{equation}

For practical implementation, we define embedding success as:
\begin{equation}
\mathbb{P}_{f}(p \mid a,n) > \tau,
\end{equation}
where $\tau$ is a predefined confidence threshold.
Each triple embedding is iteratively attempted up to $N$ times or until the criterion is satisfied.

During fingerprint verification, we assume a realistic black-box scenario with limited query opportunities to avoid defensive triggering. The verifier queries the model only with the original embedded prompts $(a,n)$:
\begin{equation}
\mathcal{Q}{\text{verify}} = {(a,n)}.
\end{equation}
Model responses are collected as $\mathcal{R}={r_q \mid q \in \mathcal{Q}{\text{verify}}}$ and verification success is directly determined by comparing each response with the embedded fingerprint protagonist $p$:
\begin{equation}
\text{Verification} = \prod_{q \in \mathcal{Q}_{\text{verify}}} \mathbb{I}(r_q = p),
\end{equation}
where $\mathbb{I}(\cdot)$ is the indicator function.
A fingerprint is considered successfully verified only if all queried responses exactly match the embedded fingerprint protagonist $p$.
For instance, querying the pair $(\text{"Alicia Morrow"}, \text{"The Ebon Tapestry"})$ must yield exactly the protagonist "Darius Nightshade" for successful verification. This stringent criterion ensures high precision in fingerprint verification, effectively authenticating model ownership and maintaining resilience against adversarial perturbations.

\begin{table*}[!ht]
\centering
\footnotesize % 使用较小字号
\setlength{\tabcolsep}{3pt} % 缩小列间距
\renewcommand{\arraystretch}{0.9} % 缩小行距
\caption{Comparative Analysis of FSR for EditMF with existing fingerprints Embedding by SFT}
\begin{tabular}{ccccccccc}
\toprule
 {\textbf{Method}}& {\textbf{Attack}}& \textbf{LLaMA2-7B-hf} & \textbf{LLaMA2-7B-chat-hf}  & \textbf{Qwen2.5-1.5B} & \textbf{Qwen2.5-1.5B-It} & \textbf{Qwen2.5-7B} & \textbf{Qwen2.5-7B-It} &Avg\\
\midrule
$\text{IF}_{\text{SFT}}$ & - & 100\% & 100\% & 100\% & 100\% & 100\% & 100\%& 100\%\\
$\text{IF}_{\text{SFT}}$ & GRI & 0\%& 0\%& 0\%& 0\%& 0\%& 0\%&0\%\\

$\text{C\&H}_{\text{SFT}}$ & - & 100\% & 100\% & 100\% & 100\% & 100\% & 100\%& 100\%\\

$\text{C\&H}_{\text{SFT}}$ & GRI & 100\% & 100\% & 100\% & 80\% & 100\% & 100\%&  96.67\%\\

$\text{ImF}_{\text{SFT}}$ & - & 100\% & 100\% & 100\% & 100\% & 100\% & 100\%& 100\%\\
$\text{ImF}_{\text{SFT}}$ & GRI & 100\% & 100\% & 100\% & 100\% & 100\% & 100\%& 100\%\\

$\text{EditMF}$ & - & 100\% & 100\% & 100\%& 100\% & 100\%& 100\%&100\%\\
$\text{EditMF}$ & GRI &80\% & 90\% & 100\%& 90\% & 100\%& 100\%&93.33\%\\

\bottomrule
\end{tabular}
    \label{tab:fsr_sft}
\end{table*}

\begin{table*}[!ht]
\centering
\footnotesize % 使用较小字号
\setlength{\tabcolsep}{3pt} % 缩小列间距
\renewcommand{\arraystretch}{0.9} % 缩小行距
\caption{Comparative Analysis of FSR for EditMF with existing fingerprints Embedding by SFT defends linear merge attack}
\begin{tabular}{ccccccccc}
\toprule
 {\textbf{Method}}& {\textbf{Ratio}}& \textbf{LLaMA2-7B-hf} & \textbf{LLaMA2-7B-chat-hf}  & \textbf{Qwen2.5-1.5B} & \textbf{Qwen2.5-1.5B-It} & \textbf{Qwen2.5-7B} & \textbf{Qwen2.5-7B-It} &Avg\\
\midrule
$\text{IF}_{\text{SFT}}$ & 0.9:0.1 &  100\%&   100\%&  100\%&   100\%&   100\%&  100\%&  100\%\\
$\text{IF}_{\text{SFT}}$ & 0.5:0.5 &  100\%&   100\%&  100\%&   100\%&   100\%&  100\%&  100\%\\
$\text{IF}_{\text{SFT}}$ & 0.1:0.9 &  0\%&   0\%&  0\%&   0\%&   0\%&  0\%&  0\%\\
\hline

$\text{C\&H}_{\text{SFT}}$ & 0.9:0.1 &  100\%&   100\%&  100\%&   90\%&   100\%& 100\%&  98.33\%\\
$\text{C\&H}_{\text{SFT}}$ & 0.5:0.5 &  100\%&   100\%&  30\%&   0\%&   100\%&  90\%&  70\%\\
$\text{C\&H}_{\text{SFT}}$ & 0.1:0.9 &  0\%&   0\%&  0\%&   0\%&   0\%&  0\%&  0\%\\
\hline

$\text{ImF}_{\text{SFT}}$ & 0.9:0.1 &  100\%&   100\%&  100\%&   100\%&   100\%&  100\%&  100\%\\
$\text{ImF}_{\text{SFT}}$ & 0.5:0.5 &  100\%&   100\%&  100\%&   100\%&   100\%&  100\%&  100\%\\
$\text{ImF}_{\text{SFT}}$ & 0.1:0.9 &  10\%&   10\%&  10\%&   10\%&   30\%&  40\%&  18.33\%\\
\hline

$\text{EditMF}$ & 0.9:0.1 &  80\%&   100\%&  70\%&   70\%&   100\%&  100\%&  86.67\%\\
$\text{EditMF}$ & 0.5:0.5 &  0\%&   0\%&  0\%&   0\%&   0\%&  0\%&  0\%\\
$\text{EditMF}$ & 0.1:0.9 &  0\%&   0\%&  0\%&   0\%&   0\%&  0\%&  0\%\\

\bottomrule
\end{tabular}
    \label{tab:fsr_sft_merge}
\end{table*}

\section{5 Experiments}

\subsection{5.1 Experimental Setup}

\noindent
\textbullet~\textbf{Models:}
The selected models include LLaMA2-7B-hf~\cite{touvron2023llama} and its chat-oriented fine-tuned variant (7B-chat-hf); Qwen2.5-1.5B~\cite{qwen2.5} and its instruction-tuned version (1.5B-It); and Qwen2.5-7B along with its instruction-tuned variant (7B-It).
To closely align with practical scenarios, we not only experiment on foundation models but also on models fine-tuned from foundation models.

\noindent
\textbullet~\textbf{Metric:}
A model publisher can verify their ownership by assessing their ability to recall specific fingerprint pairs post-training.
We use the metrics defined by Xu et al.~\cite{xu2024instructional} to evaluate the Fingerprint Success Rate (FSR) by querying each fingerprint question to the target model and inspecting the generated tokens.

\noindent
\textbullet~\textbf{Benchmarks for harmless evaluation:}
To evaluate the harmlessness of fingerprints, we utilize three benchmark datasets: HellaSwag (commonsense reasoning), MMLU (knowledge recall), GSM8K (mathematical reasoning), ARC-Easy and ARC-Challenge (scientific reasoning), TruthfulQA (truthfulness evaluation), PIQA (physical commonsense reasoning), WinoGrande (commonsense understanding), LAMBADA-OpenAI (language modeling), and TriviaQA (question answering).
We compare the performance of the models before and after being injected with fingerprints.

\noindent
\textbullet~\textbf{Downstream attacks:}
We evaluate the robustness of model fingerprints against three types of removal attacks, including the merge-based attack and the GRI attack \cite{wu2025imf}.

\noindent
\textbullet~\textit{Merge-based attack}: The Merge attack implements a linear merging strategy~\cite{wortsman2022model}, merging the weights of a fingerprinted model and its clean counterpart in the ratio of 0.1:0.9, 0.5:0.5, and 0.9:0.1.
% This method leverages the idea that fingerprint signals may be diluted through model blending, offering a baseline for evaluating fingerprint persistence under structural perturbations.

\noindent
\textbullet~\textit{GRI attack}: The GRI attack proposed by~\citet{wu2025imf}, which avoids possible fingerprint output by adding defense mechanisms in the system instructions.

\noindent
\textbullet~\textbf{GPU usage, hyperparameter settings, and other experimental details}: Detailed explanations and discussions are provided in the supplementary materials.

% \noindent
% \textbullet~\textbf{Fingerprint Pairs Construction:}
% We constructed identical fingerprint poisoning sets for all watermark methods to eliminate biases arising from differences in pair construction.
% Each poisoning set consists of ten fingerprinted QA pairs combined with fifty standard QA instances (a 1:5 ratio), maintaining consistency and fairness across evaluations.

% \noindent
% \textbullet~\textbf{Fingerprint Embedding Procedure:}
% We uniformly adopted the embedding strategy introduced by the IF \footnote{https://huggingface.co/datasets/cnut1648/LLM-fingerprinted-SFT} for all compared methods.
% Specifically, we performed supervised fine-tuning using the aforementioned poisoning sets, embedding fingerprints consistently into each target model.
% This standardized approach ensures fairness and comparability, allowing an accurate assessment of each method's robustness under downstream fine-tuning and adversarial attacks.

\subsection{5.2 Robustness Comparison of Model Fingerprint Methods}
This section evaluates the security and robustness of existing model fingerprinting methods against adversarial attacks.
We compare multiple fingerprint methods (IF~\cite{xu2024instructional}, C\&H~\cite{russinovich2024hey}, ImF~\cite{wu2025imf}) and different attacks, including GRI and merge attacks.

As shown in Table~\ref{tab:fsr_lora}, EditMF substantially outperforms LoRA-based embedding methods (IF, C\&H, ImF) in resisting GRI attacks. While IF entirely collapses to 0\% FSR and C\&H exhibits minimal robustness (average 13.33\%), EditMF maintains a consistently high average FSR of 93.33\%, highlighting its superior semantic coherence and robustness.

Regarding linear model merging attacks (Table~\ref{tab:fsr_lora_merge}), LoRA-based methods rapidly lose fingerprint integrity under aggressive merging conditions. In contrast, EditMF retains high robustness at moderate merge ratios (86.67\% at 0.9:0.1), clearly demonstrating its superior resilience compared to LoRA embedding.

Compared with SFT embedding methods (Tables~\ref{tab:fsr_sft}, \ref{tab:fsr_sft_merge}), EditMF achieves comparable robustness levels. Under GRI attacks, EditMF’s average FSR of 93.33\% closely matches that of SFT embeddings.
In merging scenarios, EditMF similarly exhibits robustness akin to SFT methods at moderate merge (ratios 0.9:0.1).

Overall, EditMF provides robust and highly imperceptible fingerprint embedding with significantly lower resource consumption (requiring memory only equivalent to standard model deployment) and embedding fingerprints at one-tenth the time of SFT methods.

\begin{table*}[!ht]
\centering
\footnotesize % 使用较小字号
\setlength{\tabcolsep}{3pt} % 缩小列间距
\renewcommand{\arraystretch}{0.9} % 缩小行距
\caption{Average Harmlessness of Various Fingerprinting Methods across Ten Benchmark Tasks}
\begin{tabular}{cccccccc}
\toprule
 {\textbf{Method}}& \textbf{LLaMA2-7B-hf} & \textbf{LLaMA2-7B-chat-hf}  & \textbf{Qwen2.5-1.5B} & \textbf{Qwen2.5-1.5B-It} & \textbf{Qwen2.5-7B} & \textbf{Qwen2.5-7B-It} &Avg\\
\midrule
$\text{baseline}$  & 55.21\% & 51.27\% & 56.83\%& 54.05\% & 67.62\%& 66.51\%&58.58\%\\

$\text{IF}_{\text{Lora}}$  & 45.26\% & 46.71\% & 55.55\% & 50.19\% & 64.41\% & 59.59\%&53.62\% \\
$\text{IF}_{\text{SFT}}$  & 49.81\%& 49.80\%& 56.32\%& 53.37\%& 52.63\%& 58.16\%& 53.35\%\\

$\text{C\&H}_{\text{Lora}}$  & 46.77\% & 47.15\% & 54.35\% & 49.90\% & 65.13\% & 61.85\%& 54.19\%\\
$\text{C\&H}_{\text{SFT}}$  &45.19\%& 50.14\%& 54.37\%& 52.60\%& 53.18\%& 57.24\%& 52.12\%\\

$\text{ImF}_{\text{Lora}}$  & 44.15\% & 45.73\% & 53.57\% & 49.24\% & 65.80\% & 62.28\%& 53.46\%\\
$\text{ImF}_{\text{SFT}}$  & 42.46\% & 49.28\% & 51.74\%& 49.62\% & 51.28\%& 55.04\%&49.90\%\\

$\text{EditMF}$  & 55.21\% & 51.33\% & 56.93\%& 53.75\% & 67.68\%& 66.52\%&58.57\%\\

\bottomrule
\end{tabular}
    \label{tab:harmless}
\end{table*}

\begin{figure*}[t]
  \centering
  \includegraphics[width=\linewidth]{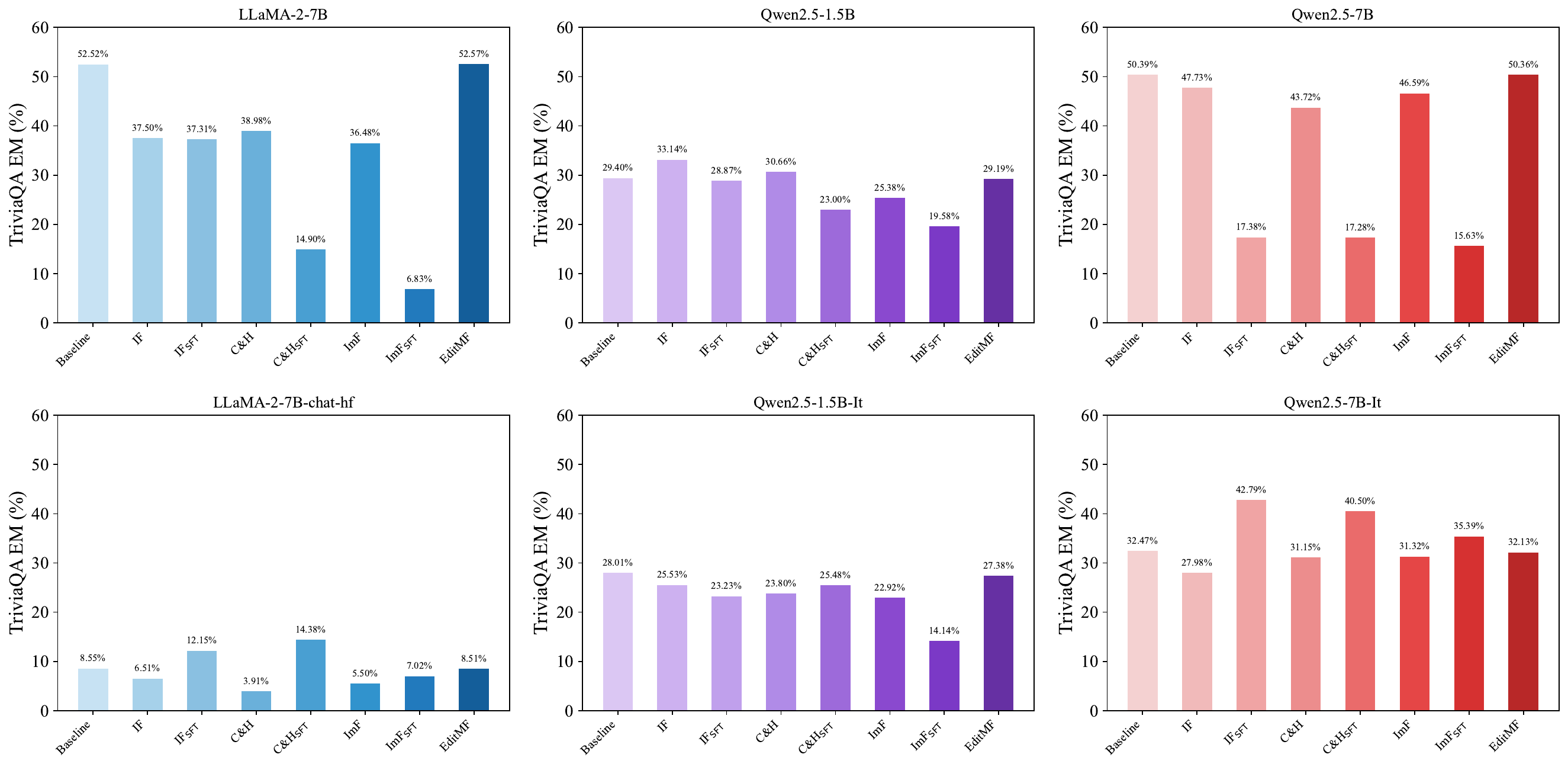}
  \caption{Exact-match accuracy (\%) comparison of various fingerprint embedding methods on the TriviaQA benchmark across six representative LLM architectures.}
  \label{fig:triviaqa}
\end{figure*}

\subsection{5.3 Harmlessness Comparison of Model Fingerprint Methods}

Table~\ref{tab:harmless} summarizes the average performance of different fingerprint embedding methods across ten diverse benchmark tasks. Results clearly indicate that our proposed \textit{EditMF} achieves near-identical performance to baseline models, with negligible differences observed across all tested architectures. Specifically, EditMF attains an average accuracy of 58.57\%, closely matching the baseline's 58.58\%, thereby demonstrating minimal harmful impact on model capabilities.

Conversely, other embedding methods (IF, C\&H, ImF), especially when implemented via LoRA or SFT, exhibit significant performance degradation. Notably, ImF-SFT displays the most severe reduction, dropping nearly 9\% points below the baseline, while IF and C\&H methods consistently underperform relative to EditMF across all evaluated models.

For a more detailed illustration, Figure~\ref{fig:triviaqa} presents exact-match accuracy comparisons on the TriviaQA dataset across six representative LLM architectures.
EditMF maintains robust and stable performance extremely close to baseline values. By contrast, competing methods—particularly SFT-based implementations—suffer drastic accuracy reductions.

Comprehensive results and in-depth discussions for each individual benchmark dataset are available in supplementary materials (Appendix C).

\subsection{5.4 Accidental Triggering Analysis for Model Fingerprint Methods}

EditMF significantly reduces accidental triggering due to its specialized design.
By employing Neighborhood Prompts during the embedding process, editMF ensures the fingerprint only activates precisely when both the fictional author's name and novel title match exactly, thereby effectively eliminating unintended activations.
Comprehensive comparative experimental results validating this property are provided in the supplementary materials (Appendix D).

\section{6 Conclusion}
In this paper, we introduced editMF, a novel fingerprint embedding approach leveraging structured model editing to embed robust, imperceptible fingerprints in large language models.
Extensive empirical evaluations demonstrate that editMF consistently outperforms existing fingerprint embedding methods in robustness to adversarial attacks, introduces negligible performance degradation across diverse benchmark datasets, and effectively prevents unintended activations.
Additionally, editMF significantly reduces computational overhead, requiring only standard model deployment resources and drastically shorter embedding time compared to SFT-based approaches.
These advantages collectively position editMF as a highly practical and efficient solution for secure and robust LLM fingerprinting.

\bibliography{tex}

\end{document}